\documentclass[unsortedadress,preprint,secnumarabic,nobibnotes,prd,showpacs]{revtex4}
\usepackage{amsfonts}
\usepackage{amsmath}
\usepackage{amssymb}

\usepackage{graphicx}%
\begin{document}

\title{Non-Abelian Born-Infeld theory without the square root}

\author{O. Obreg\'on}
\email{octavio@fisica.ugto.mx}
\affiliation{Instituto de F\'{\i}sica de la Universidad de Guanajuato, P.O. Box E-143,
37150 Le\'on Gto., M\'exico}

\date{\today}

\begin{abstract}
A non-Abelian Born-Infeld theory is presented. The square root structure that
characterizes the Dirac-Born-Infeld (DBI) action does not appear. The
procedure is based on an Abelian theory proposed by Erwin Schr\"{o}dinger
that, as he showed, is equivalent to Born-Infeld theory. We briefly mention
other possible similar proposals. Our results could be of interest in
connection with string theory and possible extensions of well known physical
results in the usual Born-Infeld Abelian case.

\end{abstract}
\maketitle

Seventy years ago Erwin Schr\"{o}dinger wrote a paper entitled Contributions
to Born's New Theory of the Electomagnetic Field \cite{schrodinger}. As is
known and he himself pointed out the classical (Dirac-Born-Infeld-DBI) Born's
theory \cite{born,born2,borninfeld,pryce,weiss} can be constructed by means of
the two vectors $\mathbf{B}$ and $\mathbf{E}$, the magnetic induction and the
electric field-strength respectively. The partial derivatives of the
Lagrangian with respect to the components of $\mathbf{B}$ and $\mathbf{E}$
define a second pair of vectors correspondingly $\mathbf{H}$, the magnetic
field and $- ~ \mathbf{D}$, the electric displacement. It was already shown by
Born that one can choose four different ways to write a Lagrange function in
terms of one of the magnetic vectors with one of the electric vectors. For
each of these theories, the Lagrangians have essentially the same structure.

\bigskip

Schr\"{o}dinger proposed a theory whose structure is entirely different from
the above mentioned. He used two complex combinations of $\mathbf{B}$,
$\mathbf{E}$, $\mathbf{H}$ and $\mathbf{D}$ as independent variables%
\begin{equation}
\mathbf{\Omega}=\mathbf{B}-i\mathbf{D}~,~\mathbf{\Sigma}=\mathbf{E}%
+i\mathbf{H}, \label{1}%
\end{equation}
and constructed a Lagrangian in such a way that the complex conjugate of one
of these variables is identical with the partial derivative of the Lagrangian
with respect to the other one. This he called the condition of conjugateness.
The Lagrangian is%
\begin{equation}
\mathcal{L}=\frac{\mathbf{\Omega}^{2}-\mathbf{\Sigma}^{2}}{\mathbf{\Omega
}\cdot\mathbf{\Sigma}}~~. \label{2}%
\end{equation}
In this Lagrangian the square root structure typical of the Dirac-Born-Infeld
action has disappeared. The Lagrangian results rational and homogeneous of the
zeroth degree.

\bigskip

Schr\"{o}dinger showed that the classical treatment of the
Lagrangian (\ref{2}) is entirely equivalent to Born's theory
(DBI), this will be shown below. He then writes {\it consequently
it can not provide us with any new insight, which could not,
virtually, be derived from Born's original treatment as well}. He
recognized that for practical calculations, however, the imaginary
vectors structure will hardly be useful. Quoting, once more
Schr\"{o}dinger {\it yet for certain theoretical considerations of
a general kind I am inclined to consider the present treatment as
the standard form on account of its extremal simplicity, the
Lagrangian being simply the $\mathit{ratio}$ of the two
invariants, whereas in Maxwell's theory it was equal to one of
them}.

\bigskip

In this work we will present a generalization of Schr\"{o}dinger's
Lagrangian (\ref{2}) to a non-Abelian gauge theory. The
complexification of the fields will provide us with a direct clue
to find the non-Abelian framework based on the Abelian
Schr\"{o}dinger's representation. As already recognized by
Schr\"{o}dinger himself, also our non-Abelian complex formalism
provides a structure hardly to handle with for actual
calculations. We will not attempt, in this work, to make
calculations with our complex non-Abelian gauge theory. The work,
however, is motivated by the fact that in the framework of string
theory, the possibility to define a non-Abelian generalization of
the standard Dirac-Born-Infeld bosonic and/or supersymmetric
actions \cite{schwarz} has been extensively explored beginning in
1990 \cite{argyres}. If this theory could be constructed, it
should represent the world-volume U(N) gauge theory that arises
when one has N coincident type II D$p$ branes. The symmetrized
trace prescription proposed by Tseytlin \cite{tseytlin} seems to
be correct up to terms of the order $F^{4}$, but if fails at
higher orders \cite{taylor}. Also in the supersymmetric set up
certain terms cannot be expressed in terms of symmetrized traces
\cite{bilal}.

\bigskip

At this stage we want only to show a consistent and relatively straightforward
way to generalize Schr\"{o}dinger's representation of the DBI action to
non-Abelian gauge field theories. It will be, as in the Abelian case, the
ratio of two invariants and does not have the usual form of a square root. The
formal structure is simple and provides us with a different starting point to
investigate another non-Abelian generalization of the Dirac-Born-Infeld
action. We will make a brief comment on other possible ways to generalize
Schr\"{o}dinger's construction.

\bigskip

We will begin by reviewing the main aspects of Schr\"{o}dinger's proposal. His
Lagrangian is then written in terms of the usual tensorial formulation in
Electrodynamics. Next we present our non-Abelian proposal, for which, as in
the Abelian case the corresponding condition of conjugateness will allow us to
identify complex fields and reduce them to the number of the usual physical
fields of the corresponding non-Abelian gauge theory. We conclude with a few remarks.

As already mentioned in the introduction Schr\"{o}dinger's proposal begins by
postulating the Lagrangian (\ref{2}) (the singular case $\mathbf{\Omega}%
\cdot\mathbf{\Sigma}=0$ is discussed in \cite{schrodinger}). The complex
combinations (1) are considered as independent variables but such that their
complex conjugates denoted by $\ast$ fulfill%
\begin{align}
\mathbf{\Omega}^{\ast}  &  =\frac{\partial\mathcal{L}}{\partial\mathbf{\Sigma
}}=-\frac{2\mathbf{\Sigma}}{\mathbf{\Omega}\cdot\mathbf{\Sigma}}%
-\frac{\mathbf{\Omega}^{2}-\mathbf{\Sigma}^{2}}{(\mathbf{\Omega}%
\cdot\mathbf{\Sigma})^{2}}\mathbf{\Omega},\label{3}\\
\mathbf{\Sigma}^{\ast}  &  =\frac{\partial\mathcal{L}}{\partial\mathbf{\Omega
}}=\frac{2\mathbf{\Omega}}{\mathbf{\Omega}\cdot\mathbf{\Sigma}}-\frac
{\mathbf{\Omega}^{2}-\mathbf{\Sigma}^{2}}{(\mathbf{\Omega}\cdot\mathbf{\Sigma
})^{2}}\mathbf{\Sigma},\nonumber
\end{align}
this he called the condition of conjugateness.

Schr\"{o}dinger then remarks that to get the field-equations corresponding to
(\ref{2}) one should not pay attention to the relation (\ref{1}), but actually
consider $\mathbf{\Omega}$ and $\mathbf{\Sigma}$ as fundamental variables. He
then, assumes (as in Born's theory and, as well known, in Maxwell's theory)
that the six complex vector $\mathbf{\Omega}$ and $\mathbf{\Sigma}$ is the
four dimensional curl of a potential four-vector, and consequently only its
four components are to be varied independently. This is equivalent to assume
that the field equations are%
\begin{equation}
\nabla\times\mathbf{\Sigma}+\frac{\partial\mathbf{\Omega}}{\partial
t}=0,~~~\nabla\cdot\mathbf{\Omega}=0. \label{4}%
\end{equation}
Then, using the conjugateness condition one can obtain by variation in the
usual way%
\begin{equation}
\nabla\times\mathbf{\Sigma}^{\ast}+\frac{\partial\mathbf{\Omega}^{\ast}%
}{\partial t}=0,~~~\nabla\cdot\mathbf{\Omega}^{\ast}=0. \label{5}%
\end{equation}
It is also shown, using (\ref{3}), that $\mathcal{L}$ becomes purely imaginary
and is also equal to%
\begin{equation}
\mathcal{L}=-\frac{\mathbf{\Omega}^{\ast2}-\mathbf{\Sigma}^{\ast2}%
}{\mathbf{\Omega}^{\ast}\cdot\mathbf{\Sigma}^{\ast}}~~. \label{6}%
\end{equation}

The stress energy momentum tensor is calculated and it is proved that there
always exists a Lorentz frame in which all the four composing three vectors
are parallel in a certain world point. Further simplification is obtained by
making use of the fact that the six components of $\mathbf{\Omega}$ and
$\mathbf{\Sigma}$ can be multiplied by a factor $e^{i\gamma}$, $\gamma$ real.
It is called the $\gamma$ -transformation. It does not interfere with the
conjugateness condition, for in (\ref{3}) the right-hand sides take the factor
$e^{-i\gamma}$. The numerical values of the Lagrangian (\ref{2}) remain
unmodified as well as those of the stress-energy-momentum tensor components.
The application to (\ref{4}) and (\ref{5}) with $\gamma= const.$ produces
another solution, though with the same energy, momentum and stress densities
as before in every world point.

A consequence of (\ref{3}) is%
\begin{equation}
\mathbf{\Omega}^{\ast}\cdot\mathbf{\Sigma}+\mathbf{\Sigma}^{\ast}%
\cdot\mathbf{\Omega}=0. \label{7}%
\end{equation}
Making use of the above mentioned Lorentz transformation one can make all
components vanish except, say, $\Omega_{1}$ and $\Sigma_{1}$ and choose
$\gamma$ so as to make $\Omega_{1}$ real. Through the relation (\ref{7}) one
can write%
\begin{equation}
\Omega_{1}(\Sigma_{1}+\Sigma_{1}^{\ast})=0, \label{8}%
\end{equation}
$\Sigma_{1}$ results imaginary and can be put as%
\begin{equation}
\Sigma_{1}=iA\Omega_{1}, \label{9}%
\end{equation}
where $A$ is a real constant. By substitution in (\ref{3}) it is easily seen
that the only allowed expressions for $\Omega_{1}$ and $\Sigma_{1}$ are%
\begin{equation}
\Omega_{1}=\frac{\sqrt{1-A^{2}}}{A},~~~\Sigma_{1}=i\sqrt{1-A^{2}}, \label{10}%
\end{equation}
$A$ takes values from $-1$ to $+1$ and the positive sign of the
square root should be taken. This is called the "standard field".
It is purely magnetic field with permeability $A^{-1}$, a purely
electric field with dielectric constant $A^{-1}$ can be obtained
by a $\gamma-$ transformation. This standard field does not
require then a further $\gamma$-transformation, but only a Lorentz
transformation would be necessary to obtain the most general
field.
The Lagrangian for the standard case results then in%
\begin{equation}
\mathcal{L}=-i\frac{1+A^{2}}{A}~~. \label{11}%
\end{equation}

The identity with Born's theory (DBI) is not fully performed in
Schr\"{o}dinger's work, it is mentioned that the condition of conjugateness
(\ref{3}) is equivalent to relations (\ref{12}) (see below). This can
easily be corroborated. The rest of the procedure is, according with the
footnote in page 472, as follows; he refers us to Born-Infeld paper \cite{pryce}
and makes two corrections to misprints, there%
\begin{equation}
\mathbf{H}=\frac{\partial L}{\partial \mathbf{B}}=
\frac{\mathbf{B}-G\mathbf{E}}{\sqrt{1+F-G^{2}}},\qquad \mathbf{D}=-\frac
{\partial L}{\partial \mathbf{E}}=\frac{\mathbf{E}+G\mathbf{B}}{\sqrt{1+F-G^{2}}}, \label{12}%
\end{equation}
with%
\begin{equation}
L=\sqrt{1+F-G^{2}}-1,\qquad F=\mathbf{B}^{2}-\mathbf{E}^{2},~~~G=\mathbf{B}\cdot\mathbf{E}. \label{13}%
\end{equation}
One then chooses a frame with $\mathbf{B}||\mathbf{E}$. Consequently $\mathbf{H}||\mathbf{D}||\mathbf{B}||\mathbf{E}$.
 By inserting
(\ref{13}) into (\ref{12}) one gets%
\begin{equation}
\mathbf{H}=A\mathbf{B},\qquad\mathbf{E}=A\mathbf{D}, \label{14}%
\end{equation}
with%
\begin{equation}
A=\sqrt{\frac{1-\mathbf{E}^{2}}{1+\mathbf{B}^{2}}}~~, \label{15}%
\end{equation}
$A$ results to be the dielectric constant and also the permeability.
Expressing $\mathbf{B}^{2}$ in terms of $\mathbf{E}^{2}$ from (\ref{15}) one
gets%
\begin{equation}
\mathbf{B}^{2}+\mathbf{D}^{2}=\frac{1-A^{2}}{A^{2}}, \label{16}%
\end{equation}
and, of course%
\begin{equation}
\mathbf{H}^{2}+\mathbf{E}^{2}=1-A^{2}. \label{17}%
\end{equation}
These last two equations reduce to Eqs. (\ref{10}) when $\mathbf{D}$ and
$\mathbf{E}$ are abolished by a $\gamma$-transformation.

To obtain Schr\"{o}dinger's Lagrangian in tensorial notation we write
explicitely the dual tensor of the electromagnetic field-strength%
\[
\tilde{F}^{\mu\nu}=\frac{1}{2}\varepsilon^{\mu\nu\gamma\delta}F_{\gamma\delta
}=\left(
\begin{array}
[c]{cccc}%
0 & -B_{1} & -B_{2} & -B_{3}\\
B_{1} & 0 & E_{3} & -E_{2}\\
B_{2} & -E_{3} & 0 & E_{1}\\
B_{3} & E_{2} & -E_{1} & 0
\end{array}
\right)  ,
\]
and%
\[
G^{\mu\nu}=\left(
\begin{array}
[c]{cccc}%
0 & -D_{1} & -D_{2} & -D_{3}\\
D_{1} & 0 & -H_{3} & H_{2}\\
D_{2} & H_{3} & 0 & -H_{1}\\
D_{3} & -H_{2} & H_{1} & 0
\end{array}
\right)  ,
\]
where $G^{\mu\nu}$ corresponds to Maxwell Electrodynamics. For a non-linear
electrodynamics in general it is calculated by means of%
\begin{equation}
G^{\mu\nu}=2\frac{\partial L}{\partial F_{\mu\nu}}=\frac{\partial L}{\partial
I_{1}}2F^{\mu\nu}-\frac{\partial L}{\partial I_{2}}\tilde{F}^{\mu\nu},
\label{18}%
\end{equation}
where $L$ is the Lagrangian of interest and%
\begin{equation}
I_{1}=\frac{1}{2}F^{\mu\nu}F_{\mu\nu},~~I_{2}=-\frac{1}{4}F_{\mu\nu}\tilde
{F}^{\mu\nu}, \label{19}%
\end{equation}
are the two Lorentz invariants (\ref{13}). Equation (\ref{18}) provides the
constitutive equations of a theory depending of $I_{1}$ and $I_{2}$
\cite{goldin}.

We define now%
\begin{equation}
\Phi^{\mu\nu}\equiv\tilde{F}^{\mu\nu}-iG^{\mu\nu}, \label{20}%
\end{equation}
which results in terms of $\mathbf{\Omega}$ and $\mathbf{\Sigma}$ in%
\[
\Phi^{\mu\nu}=\left(
\begin{array}
[c]{cccc}%
0 & -\Omega_{1} & -\Omega_{2} & -\Omega_{3}\\
\Omega_{1} & 0 & \Sigma_{3} & -\Sigma_{2}\\
\Omega_{2} & -\Sigma_{3} & 0 & \Sigma_{1}\\
\Omega_{3} & \Sigma_{2} & -\Sigma_{1} & 0
\end{array}
\right)  .
\]

The Lagrangian (\ref{2}) can now be written as%
\begin{equation}
{I\!\!\!\!L}=\frac{\frac{1}{2}\Phi^{\mu\nu}\Phi_{\mu\nu}}{-\frac{1}{4}%
\Phi^{\mu\nu}\tilde{\Phi}_{\mu\nu}}\equiv\frac{{I\!\!\!\!I}_{~1}}%
{{I\!\!\!\!I}_{~2}}~~. \label{21}%
\end{equation}
Now, in analogy with (\ref{18}), the $\mathcal{G}$ field tensor corresponding
to the Lagrangian (\ref{21}) results in the following%
\begin{equation}
\mathcal{G}^{\alpha\beta}=\frac{1}{\mathbf{\Omega}\cdot\mathbf{\Sigma}}\left(
\begin{array}
[c]{cccc}%
0 & -2\Omega_{1}+\mathcal{L}\Sigma_{1} & -2\Omega_{2}+\mathcal{L}\Sigma_{2} &
-2\Omega_{3}+\mathcal{L}\Sigma_{3}\\
2\Omega_{1}-\mathcal{L}\Sigma_{1} & 0 & 2\Sigma_{3}+\mathcal{L}\Omega_{3} &
-2\Sigma_{2}-\mathcal{L}\Omega_{2}\\
2\Omega_{2}-\mathcal{L}\Sigma_{2} & -2\Sigma_{3}-\mathcal{L}\Omega_{3} & 0 &
2\Sigma_{1}+\mathcal{L}\Omega_{1}\\
2\Omega_{3}-\mathcal{L}\Sigma_{3} & 2\Sigma_{2}+\mathcal{L}\Omega_{2} &
-2\Sigma_{1}-\mathcal{L}\Omega_{1} & 0
\end{array}
\right),  \label{22}%
\end{equation}
where $\mathcal{L}$ is the Lagrangian defined in Eq.(\ref{2}).

By demanding%
\begin{equation}
(\tilde{\Phi}^{\alpha\beta})^{\ast}=-\mathcal{G}^{\alpha\beta}, \label{23}%
\end{equation}
one gets exactly the condition of conjugateness (\ref{3}). So we have really
rewritten Shr\"{o}dinger's proposal and the same conclusions follow from this
tensorial notation.

Taking advantage of the previous tensorial notation, we propose now for
non-Abelian theories the following Lagrangian%
\begin{equation}
\L =-\frac{\frac{1}{2}T_{r}(\Phi^{\alpha\beta}\Phi_{\alpha\beta})}{\frac{1}{4}T_{r}%
(\Phi^{\alpha\beta}\tilde{\Phi}_{\alpha\beta})}\equiv\frac{{I\!\!\!\!\!I}_{~1}%
}{{I\!\!\!\!\!I}_{~2}}~~,\label{24}%
\end{equation}
where $\Phi^{\alpha\beta}=\Phi^{\alpha\beta,a}\tau_{a}$, with $\tau_{a}$ the
generators of the gauge group. As in the Abelian formulation the square root,
which is so characteristic in the DBI theory and the non-Abelian
generalization in \cite{tseytlin}, has disappeared and the Lagrangian is
rational and of the zeroth degree. The procedure formally follows in the same
manner as in the foregoing Abelian case. One can define a tensor%
\begin{equation}
{I\!\!\!G}^{\alpha\beta,a}=\frac{\partial\L }{\partial{~I\!\!\!\!\!I}_{~1}}\frac
{\partial{~I\!\!\!\!\!I}_{~1}}{\partial\Phi_{\alpha\beta,a}}+\frac{\partial
\L }{\partial{~I\!\!\!\!\!I}_{~2}}\frac{\partial{~I\!\!\!\!\!I}_{~2}}{\partial
\Phi_{\alpha\beta,a}},\label{25}%
\end{equation}
where%
\begin{equation}
\Phi^{\alpha\beta,a}\equiv\tilde{F}^{\alpha\beta,a}-i\mathcal{G}^{\alpha\beta
,a},\label{26}%
\end{equation}
with $\tilde{F}^{\alpha\beta,a}$ the dual tensor to the usual field strength
tensor of the corresponding non-Abelian theory and $\mathcal{G}^{\alpha
\beta,a}$ the corresponding tensor to the Lagrangian defined only by the
invariant~~~${I\!\!\!\!\!I}_{~1~}$, that is the one associated with the Yang-Mills
Lagrangian under consideration and~~~${I\!\!\!\!\!I}_{~2}$  its corresponding, so
called $\theta$-term. This procedure allows us to find the constitutive
equations. They can be defined by means of the use of the tensor
${I\!\!\!G}^{\alpha\beta,a}$ in (\ref{25}). In order to get the appropiate
number of fields one needs to identify complex fields. This can be done by
imposing the condition of conjugateness in this theory, in analogy with the
Abelian case we demand \footnote{in Eq. (\ref{27}) hermitian conjugation can also considered.}%
\begin{equation}
(\tilde{\Phi}^{\alpha\beta,a})^{\ast}=-{I\!\!\!G}^{\alpha\beta,a}.\label{27}%
\end{equation}
We know also how to calculate (\ref{25}) following a similar procedure as in
the Abelian formulation.

As it was assumed by Schr\"{o}dinger himself in his Abelian proposal, we
assume also here that the field strength $\Phi^{\alpha\beta, a}$ is
constructed as usual, from a potential four vector and consequently one gets
the corresponding field equations. It is straightforward to see that, if we
construct the Lagrangian with the complex field strengths (\ref{27}) it
results oppositely equal to (\ref{24}). So, that $\L $ becomes purely
imaginary as it happens in the Abelian formulation, Eqs. (\ref{2}%
,\ref{3},\ref{6}).

The non-Abelian theory (\ref{24}) is the most natural extension of
Schr\"{o}dinger's representation of DBI action. There is no ambiguity in
ordering the product of the matrices, we take simply the trace of the action
of the non-Abelian theory under consideration and divide it by the
corresponding $\theta-term$ which is also a trace. Being the denominator a
trace, it can, by example, easily be expanded in a series which multiplies the
trace in the numerator. Other possibilities can be considered, before taking
the trace in the denominator (and numerator). One can take the matrix product
in the numerator and find the inverse matrix corresponding to the denominator.
Having to multiply these matrices, one must then give a prescription to define
the Lagrangian one would one to consider; the trace or the symmetrized trace
\cite{tseytlin} can be used. This procedure is , however, not so simple as to
take the ratio of the invariants we are used to, ${I\!\!\!\!\!I}_{~1}$ and
${I\!\!\!\!\!I}_{~2}$ in the definition of $\L $.

In this work we have presented a classical non-Abelian proposal generalizing
Born-Infeld theory. Quantum aspects will be discussed in further work. Also,
in a forthcoming paper we will search for an expansion of the Lagrangian
(\ref{24}), under the condition of conjugateness (\ref{26}) in terms of the
field strength and its corresponding non-Abelian gauge fields. As mentioned,
the complex fields are identified through the condition (\ref{26}) and one
gets the same number of fields as in the corresponding standard Yang-Mills
theory. This expansion would be a first attempt towards a possible comparison
with the results in string theory \cite{taylor}. Also, applications
considering specific Lie groups will be considered to extend results that have
been studied in the Abelian case, as classical solutions that describe
solitons and brane configurations as well as physical effects related with
electric fields \cite{gibbons} approaching limiting values.

\section*{ACKNOWLEDGMENTS}

I would like to thank H. Garc\'{\i}a-Comp\'ean, C. Ram\'{\i}rez, M. Sabido and
J. L\'opez for helpful comments on this manuscript. This work was supported in
part by CONACyT, Universidad de Guanajuato and PROMEP grants.

\end{document}